# Peculiarities of the Faraday effect in gold-nanodisk/iron-garnet heterostructures

A.N. Kuzmichev[1,2]*, D.A. Sylgacheva[1,3], M.A. Kozhaev[1,4], D.M. Krichevsky[1,3], A.I. Chernov[1], A.N. Shaposhnikov[5], V.N. Berzhansky[5], F. Freire-Fernández[6], H.J. Qin[6], E. Popova[7], N. Keller[7], S. van Dijken[6], V.I. Belotelov[1,3]

[1]*Russian Quantum Center, 45 Skolkovskoye shosse, Moscow, 121353, Russia*
[2]*Moscow Institute of Physics and Technology, Moscow Region 141701, Russia*
[3]*Lomonosov Moscow State University, Faculty of Physics, Moscow 119991 Russia*
[4]*Prokhorov General Physics Institute of the Russian Academy of Science, Moscow, 119991, Russia*
[5]*Vernadsky Crimean Federal University, 4 Vernadskogo Prospekt, Simferopol, 295007, Russia*
[6]*Department of Applied Physics, Aalto University, P.O. Box 15100, FI-00076 Aalto, Finland*
[7]*Groupe d'Etude de la Matière Condensée (GEMaC), CNRS-UVSQ, Université Paris-Saclay, 78035 Versailles, France*
*Corresponding author: al.kuzmichev93@gmail.com

In this paper, matters considering the immersion of gold nanoparticles inside a magnetic medium are investigated experimentally and theoretically. Three samples with periodic arrays of Au cylinders where studied: particles on a surface of the magnetic dielectric film, inside the magnetic film and directly under the magnetic film. The largest LSPR mediated Faraday rotation resonance enhancement takes place for the case of the nanoparticles submerged inside the magnetic film. Optimal place for nanoparticles is under the magnetic medium surface at 6 nm deep in the considered configurations. It is shown that the most influence on the Faraday rotation enhancement is produced by the magnetic properties of the medium between the nanoantennas. The experimental results are in good agreement with the numerical analysis.

**Introduction.** In the past decade, investigations of plasmonic and magneto-plasmonic nanostructures and nanoantennas in particular have attracted significant research interest [1-6]. It is mainly driven by problems of controlling light at subwavelengths scales and the necessity of coupling optical devices and electrical ones. Magnetic properties of such antennas are of importance due to the non-reciprocal behavior of light propagation [7-9] and the possibility to control the polarization and intensity of light in a broad spectral range [10]. Moreover, magnetic materials are subjects of interest to spintronics and magnonics and magneto-plasmonic nanoantennas offer a way to analyze and control their magnetic state and dynamics with light.

In metallic nanoparticles localized surface plasmon resonances (LSPRs) can be excited [11]. LSPRs are attractive due to localization of the electromagnetic field in close proximity of the nanoparticle surface, making optical spectra sensitive to the surrounding medium of the nanoparticle [12]. LSPRs are rather broad due to optical losses in a metal. Therefore, it is important to deal with metals whose absorption is relatively low. For that reason, gold or silver nanoparticles are usually used. At the same time, ferromagnetic metals are very absorptive. In order to add magnetic functionality but to keep optical losses at a moderate level one could introduce magnetic dielectric materials as the surrounding of noble metal particles. Considering that, one of the most optimal configurations for magnetoplasmonic nanoantennas is represented by noble nanoparticles conjugated with a bulk magnetic insulator such as Bi-substituted iron-garnet [4,5,13-16]. For this material, it has been demonstrated that LSPRs can enhance the magneto-optical Faraday effect by about 20 times.

In this paper, we investigate theoretically and experimentally how immersion of gold nanoantennas inside the magnetic dielectric film influences the LSPR-mediated resonance of the Faraday rotation. Three kinds of samples were studied where the gold nanoparticle array is placed either on the surface of the magnetic dielectric film (Fig. 1a), or inside the magnetic film (Fig. 1b) or directly under the magnetic film (Fig. 1c). We found that the Faraday rotation of samples with Au nanoantennas placed inside the magnetic film is about 1.5 times larger than that of the other two cases. Moreover, we predict that the optimal place for nanoantennas inside the magnetic medium is 6 nm under its surface (Fig. 4). Modeling shows that the magnetic properties of the medium between the nanocylinders have at least 2 times more influence on the enhancement of the Faraday effect.

**Experimental.** For the experiments, three samples were fabricated. Sample-1 is an array of gold cylinders placed on a 500 nm thick iron-garnet film of composition $Bi_{1.4}Y_{1.6}Al_{1.55}Sc_{0.2}Fe_{3.25}O_{12}$ (BiIG-a) and covered by 100 nm sapphire (Fig. 1a). Sample-2 is an array of gold nanocylinders placed on a 30 nm thick iron-garnet film (BiIG-b) of composition $Bi_3Fe_5O_{12}$ and covered by another 90 nm thick iron-garnet film (BiIG-c) of composition $Bi_{2.8}Y_{0.2}Fe_5O_{12}$ (Fig. 1b). The bottom iron-garnet film was fabricated by pulsed laser deposition on the GGG substrate, while the top one was grown by reactive magnetron sputtering. Finally, sample-3 is an array of gold nanocylinders placed on a GGG substrate and covered by a 90 nm thick iron-garnet film (BiIG-d) (Fig. 1c). The BiIG-d film has the same composition as BiIG-c. The iron-garnet film BiIG-a was grown on gadolinium gallium-garnet (GGG) by liquid phase epitaxy [17], the films BIG-c and BIG-d – by reactive ion beam sputtering, while the film BiIG-b – by pulse laser deposition.

The gold nanocylinder arrays were fabricated by an electron-beam lithography process in a cleanroom. First, nanocylinder patterns with a period (P) of 350 nm and a diameter (d) of 150 nm were defined in a PMMA/MMA resist bilayer on top of a BiIG film or GGG substrate. Next, a $h$=50 nm thick gold layer was deposited using electron-beam evaporation. Finally, the gold nanocylinders were lift-off by placing the sample in a bath of acetone. For sample-1, the gold nanostructures were covered by a 100 nm thick $Al_2O_3$ film using atomic-layer deposition.

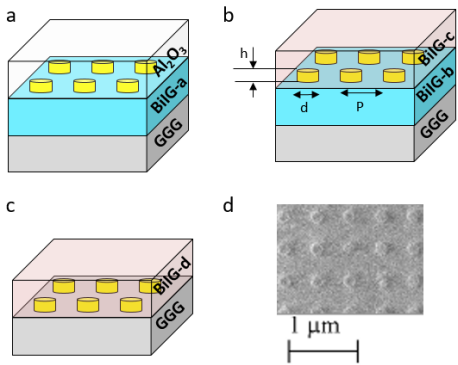

Figure 1. Schematics of sample-1 (a), sample-2 (b) and sample-3 (c). (d) SEM image of sample-3.

In order to measure the optical spectra for a wide range of wavelengths, a white light source was used (Tungsten halogen lamp, Ocean Optics HL-2000-HP). Since a point source of light is necessary for homogeneous intensity distribution within the light spot, the lamp light was focused onto a pinhole with a diameter of 100 μm. The incident light was focused in a light cone with an angle of less than 1 deg. To achieve this, a tunable collimating pinhole was placed before the focusing lens with a focal length of 150 mm. The sample was placed in-between the poles of an electromagnet (AMT&C Group Ltd, London, UK) with a 40 mm gap. The external magnetic field was controlled by a computer and a power supply. The polarization of light was adjusted using a film-based linear polarizer (Thorlabs LPVIS050) providing an extinction ratio of $10^{-5}$.

Light transmitted through the sample was collimated by an achromatic lens and focused onto the entrance slit of the spectrograph (Solar Laser Systems M266). The latter had a linear dispersion of 5.88 nm/mm. A Hamamatsu S10420-1106 2048 × 64-pixels back-thinned CCD image sensor with a spectral resolution of 0.3 nm was used.

**Results and discussion.** Optical eigenmodes of the considered structures that are excited by incident light at different wavelengths can be seen from the broadband transmittance spectra experimentally measured (Fig. 2a) and calculated (Fig. 2b). Calculations were performed by means of the Rigorous Coupled Wave Analysis (RCWA) [18] method adapted for the 2D periodic structures and utilizing permittivity in tensor form for a gyrotropic medium.

In the considered spectral range, one broad transmittance dip is observed. The calculated absolute value of the optical electric field for sample-2 indicates that this resonance is related to the excitation of a LSPR (Fig. 2 c,d). The LSPR position shifts to longer wavelengths when passing from sample-1 to sample-3. This shift is explained by an increase of the refractive index of the media surrounding the nanocylinders.

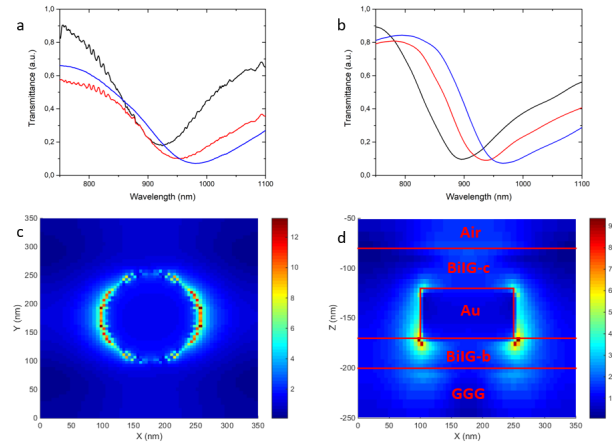

Figure 2. Measured (a) and calculated (b) transmittance spectra of three samples: black, red and blue lines correspond to the sample-1, sample-2, and sample-3, respectively. (c,d) calculated distribution of the absolute vector of the optical electric field in the vicinity of a nanocylinder in sample-2. Two different cross-sections are shown: (c) in the interface between the nanocylinder and BiIG-b film, (d) in the vertical central cross-section of the nanocylinder. The red lines mark borders of the nanoparticle and layers. The Color bar values are given in arbitrary units.

As expected, the Faraday effect is increased at the LSPR wavelength (Fig. 3). In order to evaluate the LSPR-mediated enhancement of the Faraday effect, we plot here ΔΦ - the difference between the Faraday angles measured through the part of the sample containing gold nanocylinders and through the part of the sample without nanocylinders. The largest ΔΦ is measured on sample-2, where nanocylinders are fully surrounded by the magnetic medium.

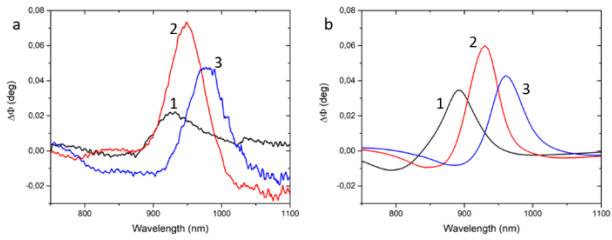

Figure 3. Measured (a) and calculated (b) difference between the Faraday angles measured through the part of the sample containing nanocylinders and through the part of the sample without nanocylinders. Black, red and blue lines correspond to the sample-1, 2, and 3, respectively.

Almost two-times smaller ΔΦ is observed for sample-3, where the nanocylinders are surrounded by the magnetic medium from the top and on the side. Finally, the lowest ΔΦ is recorded for sample-1, where the nanocylinders touch the magnetic film at the bottom only. Corresponding calculations give similar behavior.

Consequently, we conclude that the most efficient enhancement of the Faraday effect through the excitation of a LSPR is achieved by immersing the nanocylinders inside the bulk of the magnetic medium. Let us now investigate how the position of the gold nanocylinders inside the magnetic film modifies ΔΦ (Fig. 4). To do this, spectra of transmittance and ΔΦ were calculated for different immersion depths of the nanocylinders. Here, an immersion depth of 50 nm corresponds to the case where the top of the nanocylinders aligns with the upper surface of the magnetic film. The other parameters of the structure are the same as in sample-3.

The LSPR dip in transmittance shifts towards longer wavelengths since the average refractive index of the medium surrounding the nanoparticles increases for deeper immersion. The minimum value of the transmittance dip decreases gradually. On the contrary, the Faraday effect depends monotonically on the immersion depth. It can be seen that at an immersion depth of 56 nm, which corresponds to fully covered nanocylinder whose top base is 6 nm below the upper surface of the magnetic film, the Faraday rotation enhancement is the largest.

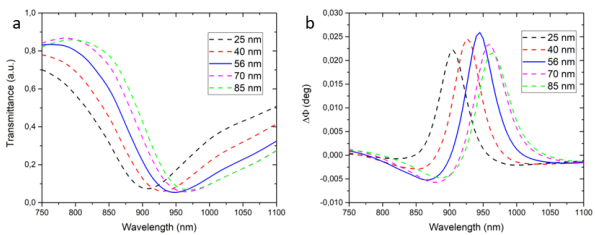

Figure 4. Calculated transmittance spectra (a), calculated Faraday rotation enhancement spectra (b). For an immersion depth of 25 nm, the resonance wavelength is 900 nm. The resonance moves towards longer wavelengths if the immersion depth is increased from 25 nm to 85 nm.

The influence of the surrounding medium on the Faraday effect mediated by the excitation of a LSPR in the nanocylinders is further analyzed by considering three refined cases: (i) a magnetic medium on top of the nanocylinders (blue region in Fig. 5a), (ii) a magnetic medium on the bottom of the nanocylinders and (iii) a magnetic medium in between the nanocylinders. We assume that other parts of the sample (white ones in Fig. 5a) are made of a dielectric with the same permittivity as the magnetic medium but without magnetization. This allows us to avoid any reflections at the interfaces between magnetized and nonmagnetized parts. Nanocylinders are located in the middle of the sample. The Faraday rotation normalized to the Faraday effect for the same samples but without nanocylinders (as is sketched in Fig. 5a,#4) are shown in Fig. 5b. The most pronounced enhancement takes place for case-3. Therefore, the Faraday effect at the LSPR is mostly sensitive to the parts of the dielectric film in between the nanocylinders. It explains the experimentally observed larger increase of the Faraday effect for the nanocylinders immersed inside the iron-garnet layer.

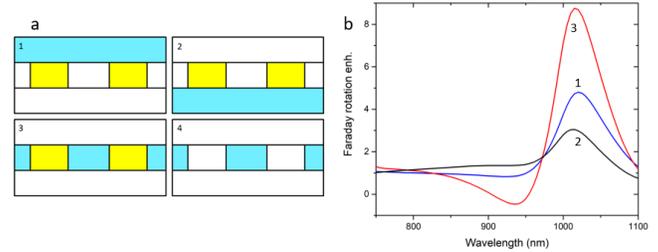

Figure 5. Structure designs for calculations (a) and corresponding Faraday rotation spectra (b). The numbers on the spectra correspond to the in (a). Blue regions are magnetized dielectric, white ones are nonmagnetized dielectric of the same permittivity, and yellow ones are gold nanodiscs.

**Conclusions.** To conclude, we considered how the position of plasmonic nanocylinders in a magnetic medium influences the Faraday effect under the excitation of localized plasmons. Our experimental data reveal that the most favorable conditions for the Faraday effect are met when the nanocylinders are places inside the magnetic dielectric. Calculations predict that the optimal situation is achieved when the nanocylinders are fully submerged into the magnetic medium staying about 6 nm below the upper surface of the magnetic film. Also, we showed that the magnetic properties of the medium between the particles have at least a 2 times greater influence on the Faraday effect than other parts of the sample.

**Funding.**, This work was financially supported by Russian Science Foundation (17-72-20260) in the part of calculation and magneto-optical experiments, by the RF Ministry of Education and Science in the framework of the state task (project no. 3.7126.2017/8.9) in the part of fabrication BIG films by reactive ion beam sputtering and by the Academy of Finland (Project Nos. 317918 and 316857) in the part of sample fabrication. Lithography of the gold nanocylinders was performed at the Micronova Nanofabrication Centre, supported by Aalto University.